\documentclass[9pt,twocolumn,english]{extarticle}

\usepackage[utf8]{inputenc}
\usepackage[T1]{fontenc}
\usepackage{babel}

\usepackage{graphicx}
\usepackage{amsfonts}
\usepackage{amsmath}
\usepackage{csquotes}
\usepackage{amssymb}
\usepackage{mathrsfs} 
\usepackage{lipsum}
\usepackage{color}
\usepackage{siunitx}
\usepackage{authblk}
\usepackage[right=2cm, left=2cm, bottom=2cm, top=2cm]{geometry}
\usepackage{mathtools, cuted}
\usepackage[dvipsnames]{xcolor}
\usepackage{sidecap}

\usepackage[font=small,labelfont=bf]{caption}

\usepackage{comment}
\newcommand{\rpb}{r_\mathrm{pb}}

\usepackage{bm}
\usepackage{amsmath}

\title{\Large \textbf{Roughness-induced friction in liquid foams}}
\author[1]{Manon Marchand}
\author[1]{Fr\'ed\'eric Restagno}
\author[1]{Emmanuelle Rio}
\author[1]{Fran\c{c}ois Boulogne}
\affil[1]{\small Universit\'e Paris-Saclay, CNRS, Laboratoire de Physique des Solides, 91405, Orsay, France}

\date{\today}
\begin{document}

\twocolumn[
    \begin{@twocolumnfalse}
        \maketitle
        \begin{abstract}
        Complex liquids flow is known to be drastically affected by the roughness condition at the interfaces.
        We combined stresses measurements and observations of the flow during the motion of different rough surfaces in dry liquid foams.
        We visually show that three distinct friction regimes exists: slippage, stick-slip motion, and anchored soap films.
        Our stress measurements are validated for slippage and anchored regimes based on existing models, and we propose a leverage rule to describe the stresses during the stick-slip regime.
        We find that the occurrence of the stick-slip or anchored regimes is controlled by the roughness factor, defined as the ratio between the size of the surface asperities and the radius of curvature of the Plateau borders.
        \end{abstract}
    \end{@twocolumnfalse}

]



The flow of complex fluids -- polymers \cite{Priestley2005, Boukany2010}, emulsions \cite{Goyon2008}, granular materials \cite{Siavoshi2006, Pouliquen1999}, bubbly liquids \cite{Germain2016}, foams \cite{Kabla2003, Golemanov2008} -- near a solid surface displays different behaviors depending on the surface properties.
Often, these systems exhibit a slippage on smooth surfaces but deform and flow on rough surfaces.
Therefore, asperities are commonly added on the measurement apparatus to suppress the interfacial dissipations \cite{Barnes1995, Cloitre2017, Cohen-Addad2013}.
Here, we focus our study on the flow of foams for which the elementary elements, the bubbles, can be observed directly.

The friction of an elongated bubble in a smooth round capillary has been studied in the pioneering work of Bretherton at low capillary numbers $\mathrm{Ca} = \mu V / \gamma$, where $\mu$ and $\gamma$ are the dynamic viscosity and air-liquid surface tension of the solution, and $V$ is the meniscus velocity in the capillary \cite{Bretherton1961}.
It is shown experimentally and theoretically that the stress follows a $\mathrm{Ca}^{2/3}$ power law.
This prediction is also valid for a single bubble in a flat cell \cite{Park1984}, and can be extended for foams \cite{Cantat2013}.
The surface rheology of the foaming solution influences the power law, but in this study we work with a solution for which this issue is negligible \cite{Cantat2013, Denkov2005}.
Also, experimental measurements provide a correction accounting for the liquid fraction of the foam $\varphi_\ell$ \cite{Raufaste2009}.

In rheology, the wall slippage of the probed material must be avoided to ensure that the dissipation is localized in the bulk of the sample.
In the case of foams, a commonly adopted rule-of-thumb to prevent slippage consists in adding asperities larger than the bubble size on the walls \cite{Khan1988}, and to assert either visually \cite{Katgert2010} or by showing that the rheological measurements do not depend on the confinement \cite{Denkov2005, Cloitre2017}, that there is no remaining slippage.
This trial and error process is not always possible and the characterization of the effect of different roughness sizes has been studied for single wet bubbles on walls \cite{Germain2016}, but not for dry liquid foams.

In this Letter, we explore systematically the effect of different sizes of wall asperities $a$ on the flow of dry liquid foams and we identify three friction regimes at zero, high or intermediate roughness. The contribution of the roughness size $a$, of the liquid fraction $\varphi_\ell$, and of the inserting velocity in the foam $V$ to the stresses are rationalized in each regime. The roughness size is shown to change the stresses over one order of magnitude. We show that the roughness factor $a/r_\mathrm{pb}$ is a single criterion for the crossover between wall and bulk dissipation regimes. Interestingly, this parameter is independent of the imposed velocity.

\begin{figure}[ht]
    \centering
    \includegraphics[width=\linewidth]{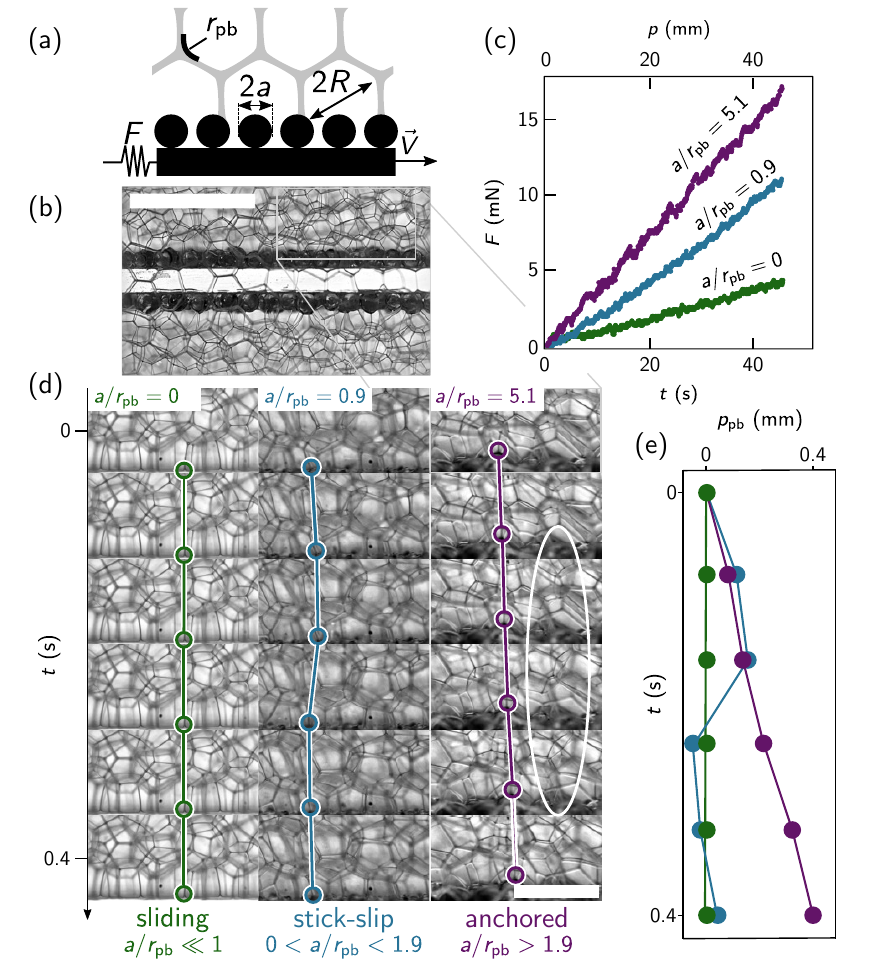}
    \caption{Foam flow dynamics near a solid surface decorated with glass beads.
    (a) Side view of the surface of roughness size $a$ moving at constant speed $V$ in a foam of bubble radius $R$ and radius of curvature of Plateau borders $r_\mathrm{pb}$.
    The force $F$ exerted by the foam on the surface is recorded.
    (b) Photograph corresponding to the schematic (a).
    The scale bar is $5$~mm.
    (c) Raw force measurements as a function of time $t$ and position of the surface $p$ for one smooth and two patterned surfaces translating at $V=1$~mm/s from left to right.
    (d) Visualization of the Plateau borders in contact with the patterned surfaces for three different $a/r_\mathrm{pb}$.
    The bottom of each picture is the surface moving at $1$~mm/s.
    We mark the position $p_\mathrm{pb}$ of a Plateau border with a circle.
    A plastic event is circled in white.
    The scale bar represents $2$~mm.
    (e) Position of the Plateau borders $p_\mathrm{pb}$ for the three friction regimes, reported from (d).
    }
    \label{fig:principle}
\end{figure}

Our experiment consists in inserting horizontally a controlled rough surface in a dry monodisperse foam generated in a transparent container (Fig.~\ref{fig:principle}a and b).
We produce the foam by blowing air through needles in a foaming solution.
The obtained bubble radius is $R = 0.70 \pm 0.05$~mm, and the liquid fraction $\varphi_\ell$ of the foam can be varied between $0.03$ and $8$~\% by changing the working height above the liquid-foam interface \cite{Maestro2013} (Fig.~S2).
For such dry foams, the radius of curvature of the Plateau borders is given by $r_\mathrm{pb} = R \sqrt{\varphi_\ell / 0.33}$ \cite{Koehler2000}.
It varies in the range $0.02-0.4$~mm in our experiment.
Model rough surfaces are obtained by gluing glass beads of mean radii $a$ on microscope glass slides (Fig.~S1), Fig. \ref{fig:principle}a).
We insert the surfaces in foams at constant velocities $V$ from $0.5$ to $20$~mm/s, corresponding to capillary numbers $\mathrm{Ca}$ in the range $1.7\times 10^{-5}$ to $6.7 \times 10^{-4}$.
Simultaneously, we measure the tangential force $F$ exerted by the foam on the surface (Fig.~S3).

%
%

\begin{figure*}[htb]
    \includegraphics[width=\linewidth]{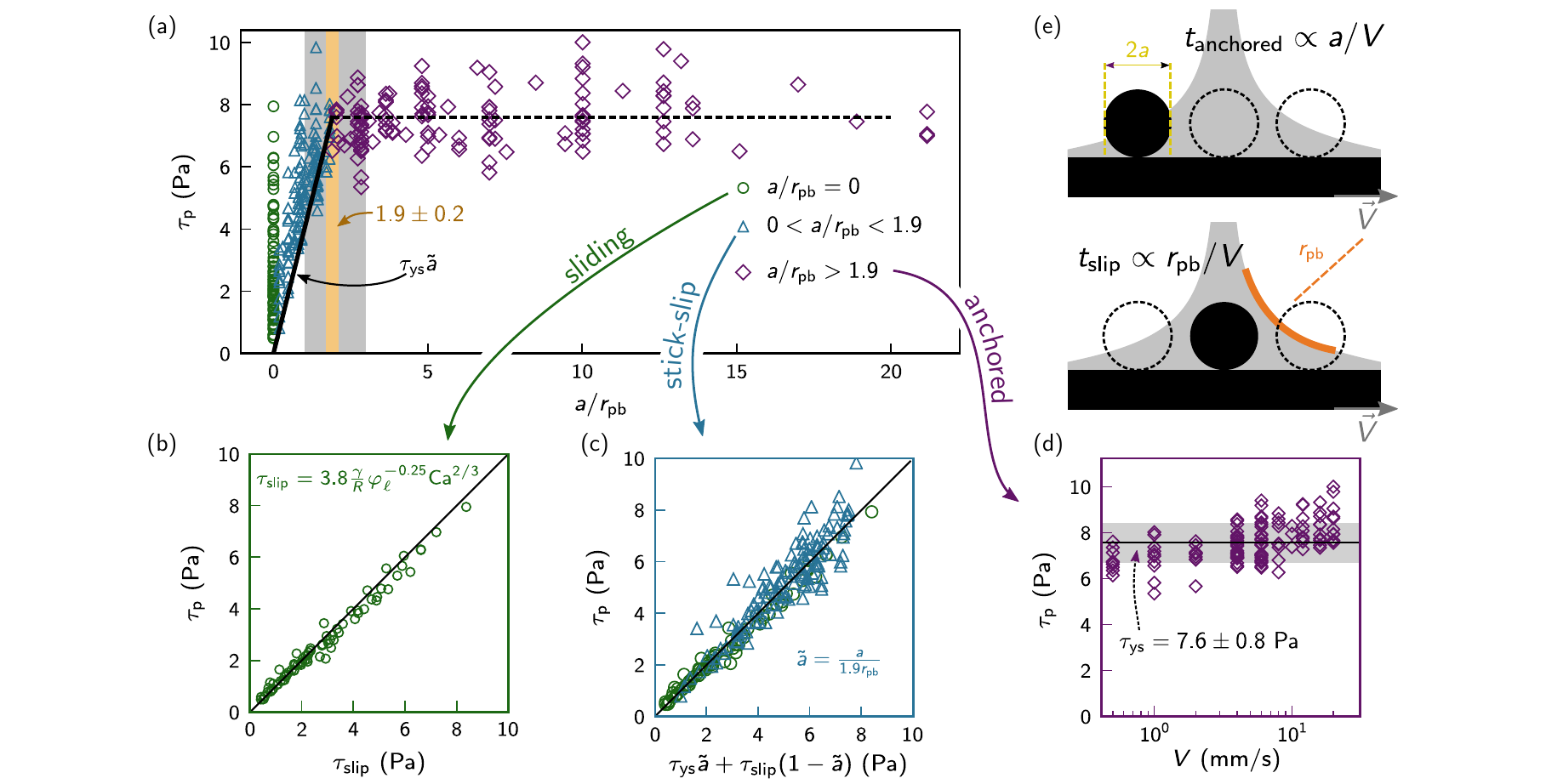}
    \centering
    \caption{Stress measurements and identification of the three friction regimes.
    (a) Stress $\tau_\mathrm{p}$ obtained from the force measurements as a function of the roughness factor $a/r_\mathrm{pb}$.
    The solid black line represents the equation (\ref{eq:stick-slip}) for $V\rightarrow 0$, and the dashed black line is $\tau_\mathrm{ys}$ with $K=2.9$.
    The gray area shows the transition between the stick-slip and the anchored regime as deduced from observation of the videos, and the orange area is the transition obtained by optimizing $k$ in Eq.~(\ref{eq:stick-slip}).
    (b) For $a/r_{\mathrm{pb}}=0$, stress measurements as a function of the equation (\ref{eq:bretherton}).
    (c) Stress values for the sliding (green circles) and the stick-slip (blue triangles) conditions, as a function of equation (\ref{eq:stick-slip}).
    (d) In the anchored regime, measured stresses as a function of the imposed velocity $V$.
    The solid line is the mean of all stresses $\tau_\mathrm{ys}$ and the gray area represents the standard deviation.
    The deviation for the higher velocities indicates a more significant contribution of the viscous term in the Hershell-Bulkley model.
    (e) Schematics of a Plateau border moving on a rough surface.
    The beads spend a time proportional to $a/V$ going through the liquid air interface, and a time proportional to $r_\mathrm{pb}/V$ in the bulk of the Plateau border.
    }
    \label{fig:measurements}
\end{figure*}

By tracking the position of the Plateau borders $p_\mathrm{pb}$ in contact with the patterned surface during the motion, we identify three distinct friction regimes (Fig.~\ref{fig:principle}d), which are discriminated by the dimensionless roughness factor $a / r_{\mathrm{pb}}$.
The emergence of this parameter can be intuited.
When a bead is smaller than the size of the Plateau border $\rpb$, it enters the Plateau border and slides inside it whereas bigger beads experience a pinning force at the liquid air interface (Fig.\ref{fig:measurements}e).
For negligible roughness sizes compared to the characteristic size of a Plateau border, $a \ll r_\mathrm{pb}$, the Plateau borders slide on the solid surface (Supplementary Movie~S1).
For $a \sim r_\mathrm{pb}$, Plateau borders are anchored on the surfaces but occasionally jump back (Supplementary Movie S2), while for $a \gtrsim r_\mathrm{pb}$ Plateau borders remain anchored and the stress is released by plastic events between bubbles in the bulk of the foam (Fig.~\ref{fig:principle}d, e, Supplementary Movie~S3).
We will respectively refer to these regimes as sliding, stick-slip, and anchored regimes in the following discussion.
For $a/ r_\mathrm{pb}$ between $1$ and $3$, the visualizations reveal that the stress is released by mixed slip or plastic events.
This provides a first estimate of the transition between the stick-slip regime and the anchored one.

Typical force measurements as a function of time $t$, or equivalently of the position of the surface $p$, are presented for each regime in Fig.~\ref{fig:principle}c.
The linear evolution of the force indicates that the probed phenomena are independent of the penetration depth of the surface, and that a constant stress value $\tau_\mathrm{p}$ can be extracted from $F(t)$ and $p(t)$ for each experiment (Fig.~S5).

%
%

We reproduce the experiment for various roughness sizes $a$,  radius of curvature of Plateau borders $r_\mathrm{pb}$, and insertion velocities $V$.
In Fig.~\ref{fig:measurements}a, we plot the stress $\tau_\mathrm{p}$ as a function of the roughness factor $a /r_\mathrm{pb}$.
The different point colors represent the three regimes identified from the images.
In this representation, the data for a single velocity collapses on a trend curve (Fig.~S4).
Complementary representations of the stick-slip and the anchored regimes presenting the influence of the asperity size $a$ and the velocity $V$ are provided in Fig.~S7.
Interestingly, tuning $a / r_\mathrm{pb}$ leads to a stress variation of about an order of magnitude, which highlights the crucial effect of the boundary conditions on stresses.
In the following, we examine each regime to rationalize the dynamics.

The sliding regime, for which $a \ll r_\mathrm{pb}$, has already been investigated \cite{Bretherton1961, Raufaste2009, Cantat2013}, and these studies validate experimentally that the stress at the solid wall can be written

\begin{equation}
    \tau_\mathrm{slip} = 3.8 \frac{\gamma}{R} \varphi_\ell^{-0.25} \mathrm{Ca}^{2/3}.
    \label{eq:bretherton}
\end{equation}
\noindent In Fig.~\ref{fig:measurements}b, the stress measurements are plotted as a function of the prediction given by equation (\ref{eq:bretherton}), which shows a good agreement and validates this experimental approach.

Now, we consider the anchored regime for the large roughness factors.
The common description of the foam rheology suggests that for shear stresses lower than a yield stress $\tau_\mathrm{ys}$, the foam behaves as an elastic solid, whereas for stresses exceeding the yield stress, the foam flows.
Phenomenologically, this behavior is described by a Hershell-Bulkley law $\tau_\mathrm{HB} = \tau_\mathrm{ys} + \eta(\dot{\gamma}) \dot{\gamma}$, where $\dot{\gamma} \sim V / \delta$ is the shear rate that scales as the velocity over a characteristic shear length $\delta$ and $\eta(\dot{\gamma})$ is the effective foam viscosity that depends on $\dot{\gamma}$ \cite{Denkov2009, Cohen-Addad2013}.
The ratio of the yield stress and the viscous term defines the Bingham number $\textrm{Bi} = \frac{\tau_\mathrm{ys}}{\eta(\dot{\gamma}) \dot{\gamma}}$ \cite{Bingham1916}.
The yield stress value is described by a semi-empirical law $\tau_\mathrm{ys} = K \frac{\gamma}{R} (\varphi_\mathrm{c} - \varphi_\ell)^2$, where $\varphi_\mathrm{c} = 0.26$ is the fraction of gaps remaining in a close-packing of hard spheres \cite{Maestro2013}, and $K$ is a proportionality factor which reported values vary between $0.5$ and $6$ \cite{Lexis2014, Rouyer2005}.
Since we are working with dry foams, we have $\varphi_\ell \ll \varphi_\mathrm{c}$, which leads to

\begin{equation}
    \tau_\mathrm{ys} = K \frac{\gamma}{R}\varphi_\mathrm{c}^2.
    \label{eq:yield_stress}
\end{equation}

In the limit of low velocities, corresponding to $\mathrm{Bi} > 1$, the contribution of the yield stress is dominant (see SI).
The Fig.~\ref{fig:measurements}d shows that the stress applied by the foam on the rough surfaces is nearly independent on the velocity and its value is predicted by the equation (\ref{eq:yield_stress}) with a prefactor $K = 2.9 \pm 0.3$, which is in agreement with previously recorded values \cite{Lexis2014, Rouyer2005}.
We notice that the measured stress increases slightly for the larger velocities, which is reminiscent of the viscous term in the Hershell-Bulkley model.

\begin{figure}[htb]
    \centering
    \includegraphics[width=\linewidth]{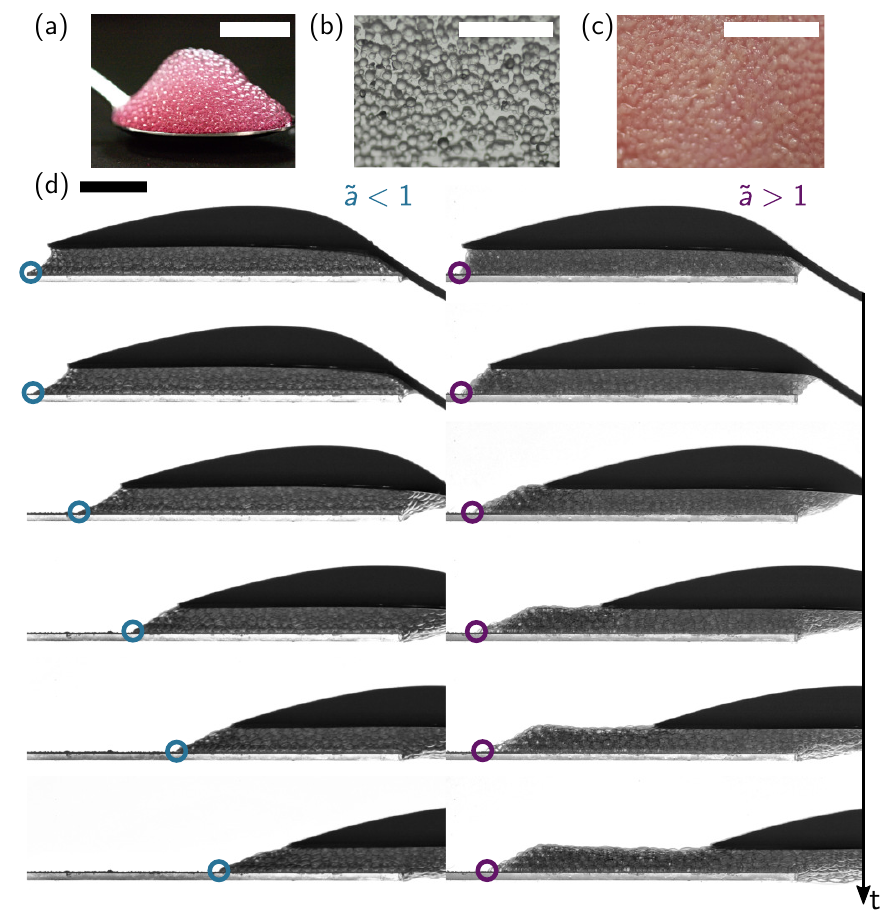}
    \caption{Edible foams and texture sensation.
    (a) Albumen foam in a spoon. The scale bar is $1$~cm. A drop of food dye is added for visualization.
    (b) Rough surface with beads of mean radius $a = 225$~$\mu$m. The scale bar is $5$~mm.
    (c) Close up picture that shows the papillae on a human tongue. The scale bar is $5$~mm.
    (d) Albumen foam sheared between a table spoon and a glass surface covered with beads of mean radius $a = 225$~$\mu$m.
    The left side is a wet foam with $\varphi_\ell \simeq 0.7$~\%, corresponding to $\tilde{a} \simeq 0.5$.
    The right side is a dryer foam with a liquid fraction $\varphi_\ell \simeq 0.02$~\%, meaning $\tilde{a} \simeq 3$.
    In those two cases the last bubble on the surface is circled to enlighten the motion of foam on the model tongue.
    There is a $1$~s interval between two consecutive images.
    The scale bar is $1$~cm.
    }
    \label{fig:application}
\end{figure}

Up to this point, we analyzed the two extreme limits of negligible and large dimensionless roughness factor $a/r_{\mathrm{pb}}$.
For roughness sizes comparable to the radius of the Plateau borders, we observe a stick-slip regime that has been observed but not characterized \cite{Buchgraber2012}, to the best of our knowledge.
Our experiments indicate that the transition between the stick-slip and the anchored regimes is continuous (Fig.~\ref{fig:measurements}a).
This suggests that the stick-slip regime is a combination of sliding, where the energy is released by viscous dissipation in the Plateau borders, and anchoring, where the energy is stored by deforming elastically the foam.

In the stick-slip regime, (Fig.~\ref{fig:measurements}e) an asperity moving in a foam goes through the liquid air interface of a Plateau border over a duration $t_\mathrm{anchored} \propto a/V$, experiencing a pinning force.
Then, the asperity moves inside the Plateau border over a typical duration $t_\mathrm{slip} \propto r_\mathrm{pb} / V$.
Therefore, we introduce the ratio of these durations, $\Tilde{a} = a/ k r_\mathrm{pb}$, where the prefactor $k$ represents the value of the transition.
This two-step description happens for each asperity on the whole probing surface.
The stick-slip events are not occurring simultaneously (Supplementary Movie S2), thus we average the individual behaviors of the Plateau borders.
Hence, we construct a leverage rule stating that the total stress is the sum of the contributions of a pinning stress for a relative duration $\Tilde{a}$ and of a viscous stress for the complementary relative duration $(1-\Tilde{a})$.
This writes

\begin{equation}
     \tau_\mathrm{stick-slip} = \tau_\mathrm{ys} \Tilde{a} + \tau_\mathrm{slip}(1-\Tilde{a}).
    \label{eq:stick-slip}
\end{equation}

To determine more precisely the value of the transition $k$, we optimized equation~(\ref{eq:stick-slip}) against this parameter and we obtained a transition at $k=1.9 \pm 0.2$ (Fig.~S6a).
This value of order unity is in agreement with our estimation of the transition from direct visualizations of the Plateau border dynamics (see the orange and gray areas in Fig.~\ref{fig:measurements}a).
As shown in Fig.~\ref{fig:measurements}c, the proposed equation~(\ref{eq:stick-slip}) is in excellent agreement with the stress measurements for the different liquid fractions and surface velocities explored in this study.
In the vicinity of small velocities, we expect $\tau_\mathrm{stick-slip} \underset{V \to 0}{\rightarrow} \tau_\mathrm{ys} \Tilde{a}$, and this is observed in Fig.~\ref{fig:measurements}a and Fig.~S6b.

%
%

Thus, we identified a practical criterion $\tilde{a}$ to discriminate between surface and bulk dissipation for the friction of a foam on a solid surface.
The anchored regime is of particular interest for rheological studies focusing on the flow of foams where a no slip condition is necessary to transfer shear to the bulk of the material.
It was already known that adding some asperities by gluing sand paper or engraving grooves on a measuring apparatus reduces the wall slippage \cite{Khan1988}, but the size of these asperities had not been tuned systematically.
Our study validates this approach, since the stress in the anchored regime are independent of the size of the asperities beyond a critical roughness factor (Fig.~\ref{fig:measurements}a).
When the grit size of the sand paper is insufficient to totally eliminate the wall slippage, the mixed surface and bulk dissipation leads to a difficult interpretation of the data \cite{Golemanov2008}.
Differences in the measured stresses are also reported when the grit size is increased \cite{Jimenez-Junca2011}.
This behavior is likely the stick-slip regime we report in Fig.~\ref{fig:measurements}c.
Our contribution allows one to choose a suitable roughness to be added on the measuring systems regarding the foam properties.
The results then assess that the obtained measurements in the anchored regime will be uncorrelated with the size of the roughness.

The stick-slip regime allows, for instance, to understand the mastication of an aerated food (Fig.~\ref{fig:application}a).
Adding air bubbles in a food product affects mouthfeel perception, improves digestibility, and aids mastication \cite{Campbell1999}.
In the case of eating disorders, such as dysphagia, more viscous foods are proven to be safer \cite{Nishinari2019}.
With viscosities typically $10^3$ times bigger than the viscosity of water \cite{Marze2005}, foams could help patients.
We make a visualization experiment to illustrate the tunability of perception during the mastication of a foamed food.
Human tongue asperities have a fixed size, therefore the remaining adjustable parameters in this case are the bubble size and the liquid fraction of the foam.
A glass slide covered with asperities of mean radius $a = 225$~$\mu$m, a size comparable to human papillae \cite{Essick2003}, can be considered as a model human tongue (Fig.~\ref{fig:application}a, b and c).
We produce two edible albumen foams with a bubble size $R=0.7$~mm and two different liquid fractions, using the same technique than soap foams, but with a solution containing $1$~g of egg white powder (Louis François) and $200$~mL of deionized water.
The two foams are sheared between a table spoon moving at $10$~mm/s and the static model tongue.
In the two cases, the last bubble is circled to illustrate the slippage of the foam (Fig.~\ref{fig:application}d).
For the wetter foam for which  $\tilde{a} \simeq 0.5 < 1$, the conditions predict a stick-slip regime and the foam moves on the surface (Supplementary Movie S4), whereas for the dryer foam, for which $\tilde{a} \simeq 3 > 1$, the foam sticks to the surface (Supplementary Movie S5).
This illustrates that the results of our controlled experiment are valid in a more general context, with different shearing geometries and surfactants.
When the normalized roughness $\tilde{a}$ varies from $0$ to $1$ in the stick-slip regime, we see in Fig.~\ref{fig:measurements}a that the stress varies between $0.8$ and $9.8$~Pa, a range of stresses detectable by a human tongue \cite{Foegeding2015, Linne2017}.
Thus, tuning the macroscopic properties of a product, namely the bubble size and the liquid-gas ratio, must lead to different sensations in mouth.

In conclusion, we reveal that the foam flow near surfaces exhibit different behaviors: sliding, anchored, and stick-slip.
By recording the stresses exerted during these regimes, we show that there exists a roughness, normalized by the curvature radius of the Plateau borders of the foam, $\tilde{a} =  a / k r_\mathrm{pb}$ beyond which the dissipation is transferred from the surface to the bulk.
This transition does not depend on the imposed velocity in the explored range.
In the stick slip regime, we propose a leverage rule describing the influence of speed, radius of curvature of Plateau borders, and liquid fraction on the stresses exerted on the walls.
We measure that these stresses vary on one order of magnitude when the normalized roughness is increased.
Therefore, we contribute to an improved description of wall roughness effect on flowing foams, which is of particular interest to design surfaces given the foam properties in industrial applications.
In contrast, when surface asperities are unchangeable, as in the case of tongue papillae, our description of friction forces allows to finely tune the properties of edible foams to get creamy or gooey mouth feelings.
This is useful for inventing new foamed food products, either for the ease of feeding for dysphagic patients, or for the gourmets' pleasure.

%
%

\paragraph{Acknowledgements}

M.M. acknowledges EDPIF for funding support.
We thank J. Sanchez for programming the computer interface of the experiment, and G. Guillier for support manufacturing the force sensor.
We thank B. Dollet and C. Raufaste for discussions.

\bibliographystyle{unsrt}
\bibliography{biblio}

\newpage
\pagestyle{empty}

\begin{figure*}[h]
    \centering
    \includegraphics[width=\textwidth]{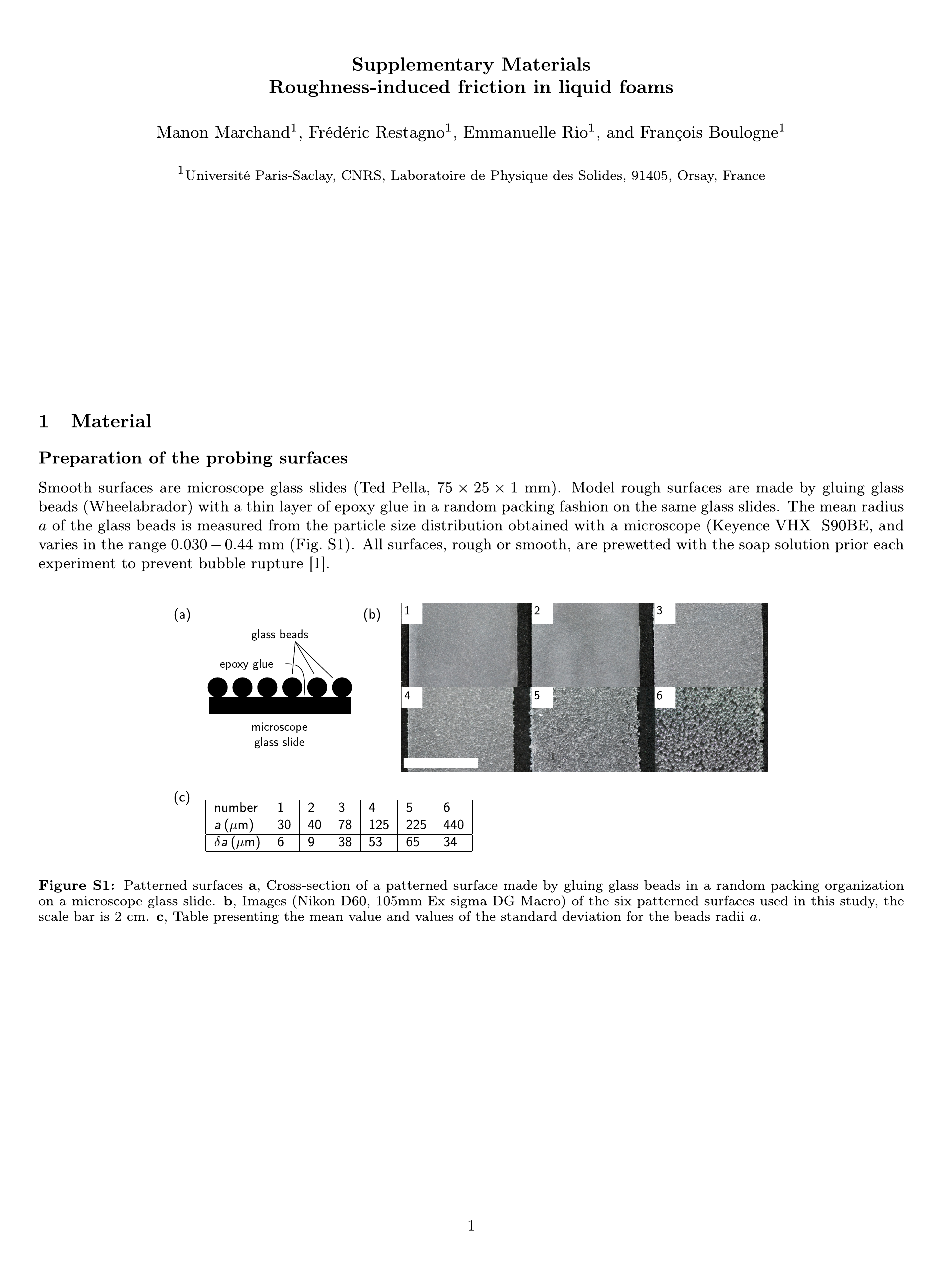}
\end{figure*}

\newpage
\pagestyle{empty}

\begin{figure*}[h]
    \centering
    \includegraphics[width=\textwidth]{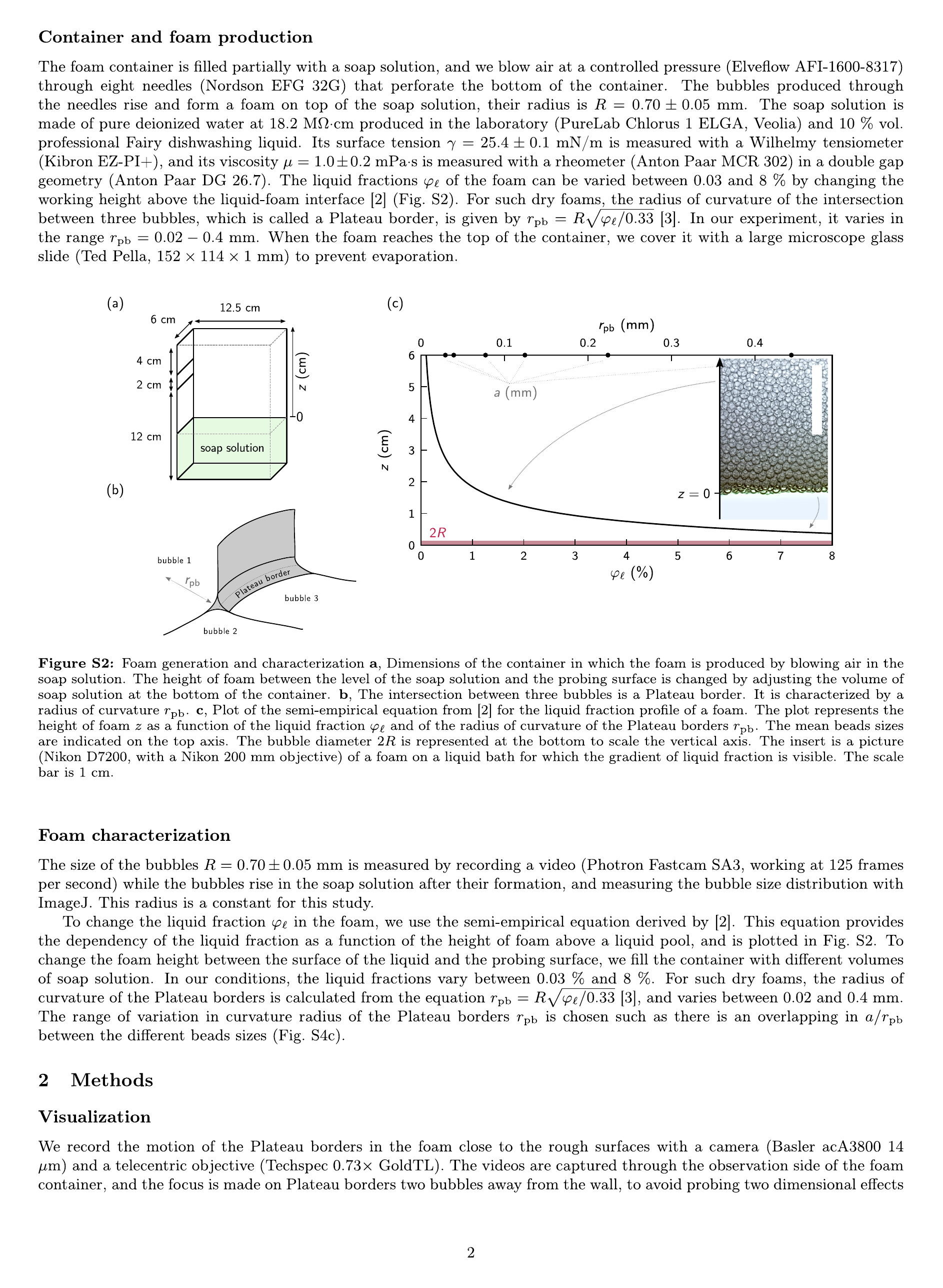}
\end{figure*}

\newpage
\pagestyle{empty}

\begin{figure*}[h]
    \centering
    \includegraphics[width=\textwidth]{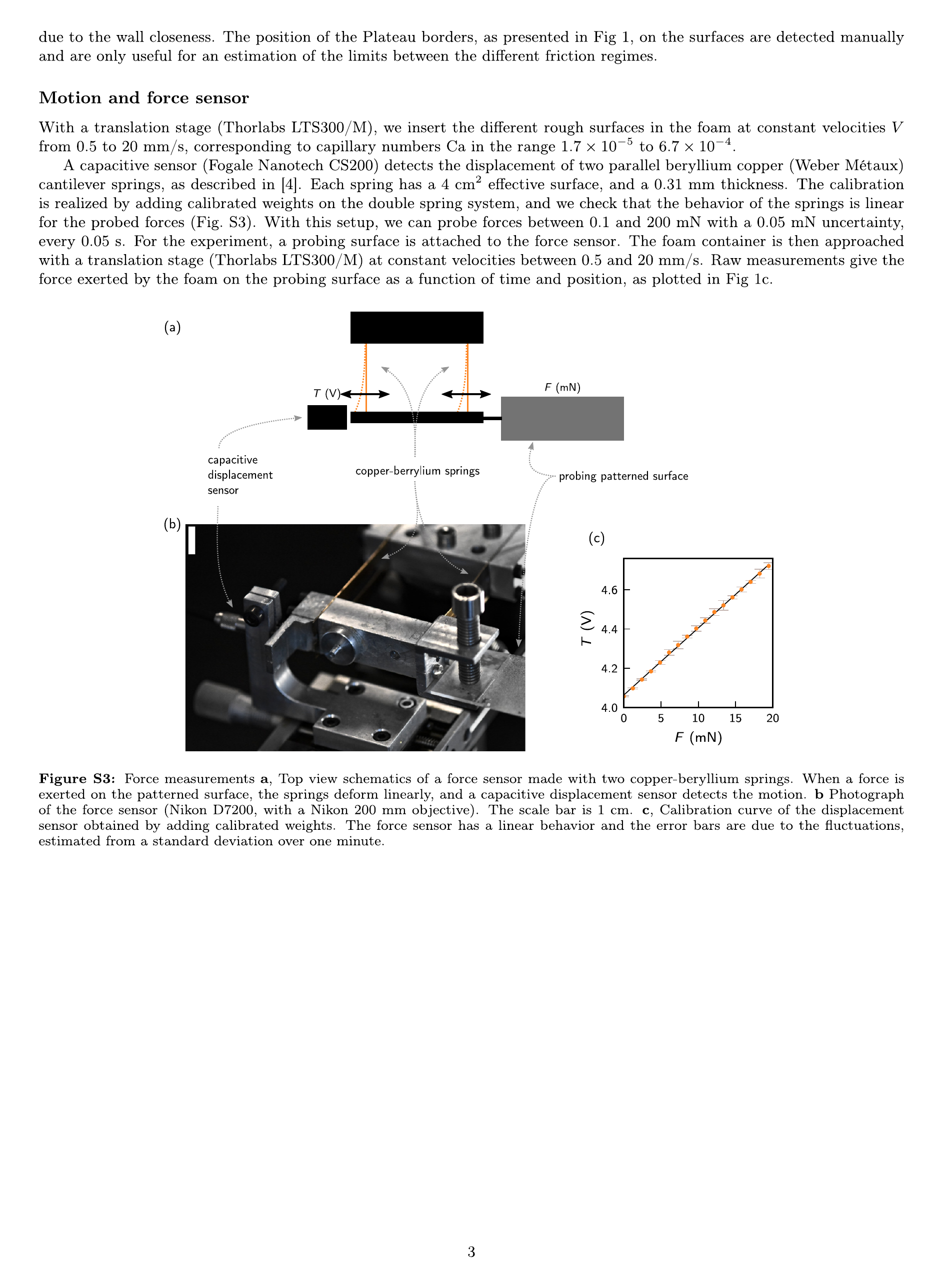}
\end{figure*}

\newpage
\pagestyle{empty}

\begin{figure*}[h]
    \centering
    \includegraphics[width=\textwidth]{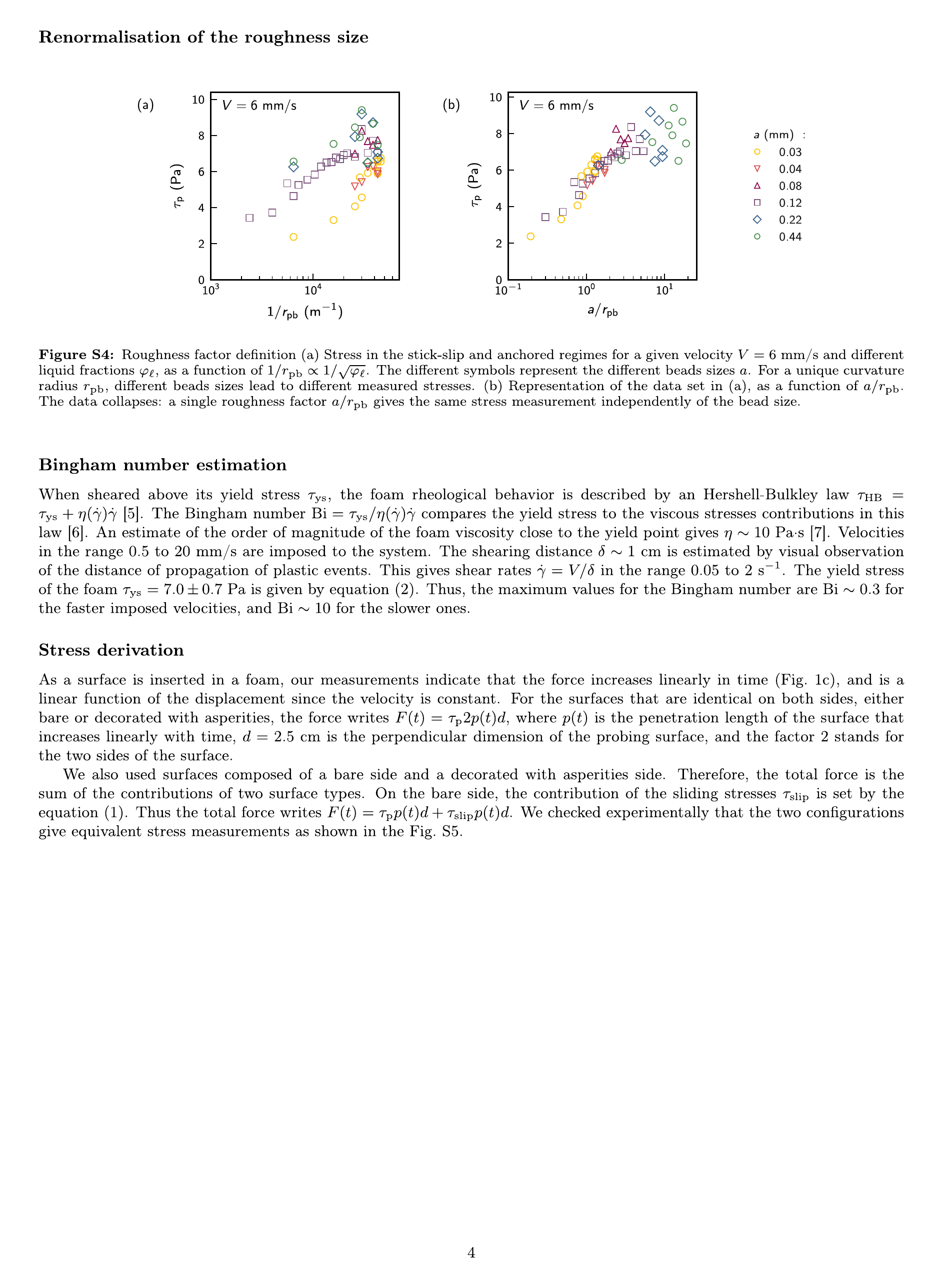}
\end{figure*}

\newpage
\pagestyle{empty}

\begin{figure*}[h]
    \centering
    \includegraphics[width=\textwidth]{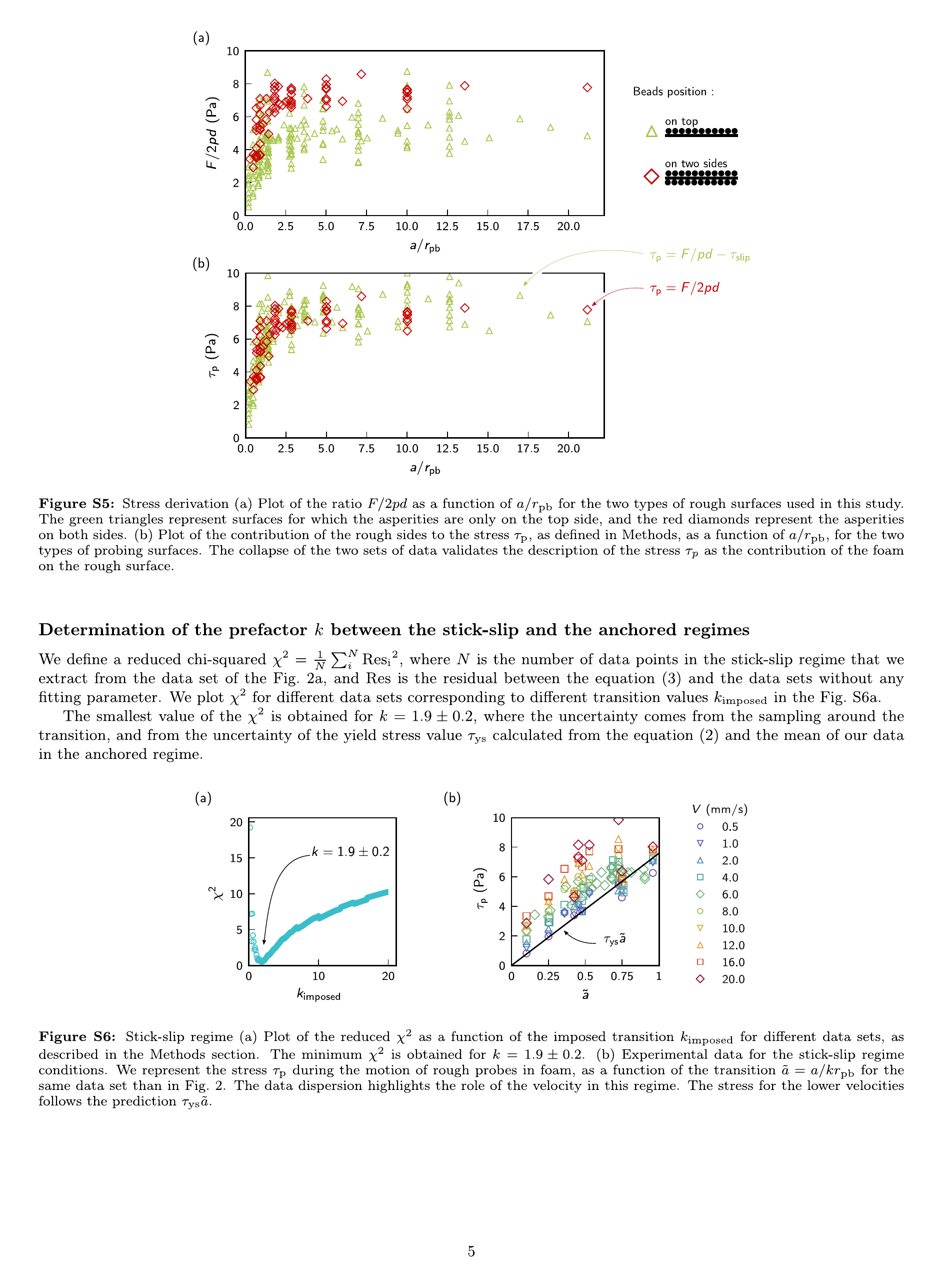}
\end{figure*}

\newpage
\pagestyle{empty}

\begin{figure*}[h]
    \centering
    \includegraphics[width=\textwidth]{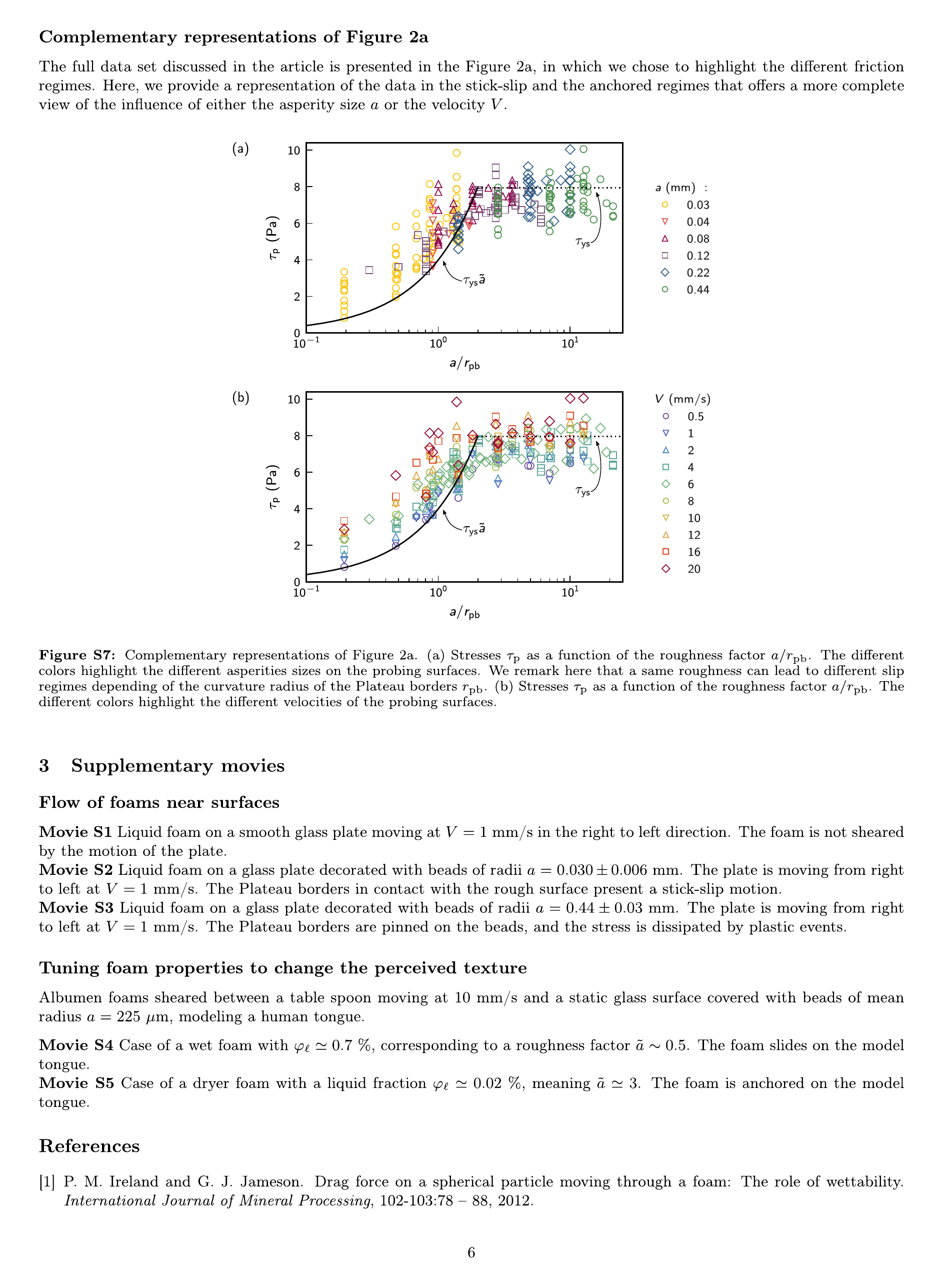}
\end{figure*}

\newpage
\pagestyle{empty}

\begin{figure*}[h]
    \centering
    \includegraphics[width=\textwidth]{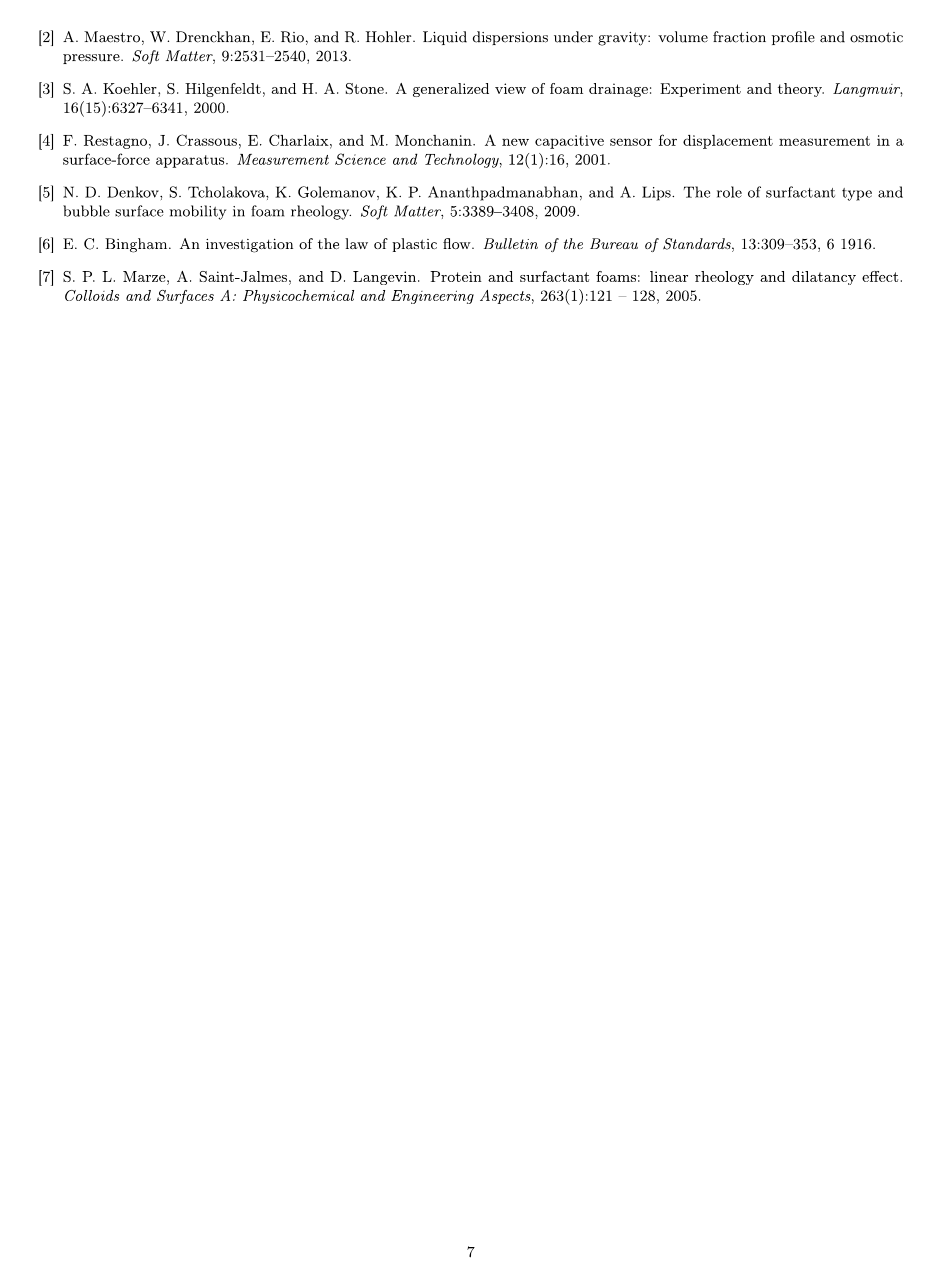}
\end{figure*}

\end{document}